# A Neutron Spin-Echo Concept for Elastic Scattering Spectroscopy (ESS) for Dynamics of Complex (Bio-) Systems

Antonio Benedetto[1,2, a)] and Gordon J. Kearley[3]

[1]*School of Physics, University College Dublin, Dublin 4, Ireland*
[2]*Laboratory for Neutron Scattering, Paul Scherrer Institut, Villigen, Switzerland.*
[3]*School of Materials Science and Engineering, UNSW Australia, Sydney, Australia.*

[a)]Corresponding author: antonio.benedetto@ucd.ie

**Abstract.** Recently, a new neutron spectroscopy for the dynamics in complex (bio-) systems has been proposed [A. Benedetto, and G. J. Kearley, Sci. Rep. **6**, 34266, (2016)]. This spectroscopy is ideal where only the overall relaxation time in a parameterless way is required, for example in complex systems, because only the elastic-scattering intensity as a function of the energy resolution is required. This has been termed "Elastic Scattering Spectroscopy" (ESS). It is based on the inflection points in the elastic-scattering intensity at the energy-resolution value corresponding to the overall system relaxation-time. A Constant wavelength (CW) option, more suitable for reactor sources, and the time-of-flight (TOF) option, more suitable for spallation sources, have already been proposed, and here we examine the concept of a third option based on neutron spin-echo (NSE), called ESS-NSE. In principle, this consists of simply measuring depolarisation at the relatively-intense elastic echo-condition, as a function of resolution, with the basic set-up being similar to standard spin-echo. In its basic set-up, ESS spin-echo can access 5 orders of magnitude in time from nanoseconds to tens of picoseconds, reaching slower relaxation processes than the CW and TOF options recently presented. However, spatial focussing is important for the small sample-sizes of biological systems, so we also explore how the fields may be shaped to enable a small neutron beam to be focussed on the sample.

## INTRODUCTION

Neutrons are an unique probe to investigate the structural and dynamical properties of panoply of different material systems ranging from solid crystals to glasses to complex fluids [1-4]. The increasing complexity of systems that are able to be studied with neutron scattering has had important implications for biological sciences. However, in some instances progress for biology (and other complex systems) has been hampered by trying to evolve the existing neutron instrumentation to accommodate biology, rather than devising specific new instrumental concepts. This is broadly analogous to the early days of using neutron scattering for engineering studies, which improved dramatically when purpose built strain-scanners arrived, rather than making the best of existing powder diffractometers.

The large incoherent scattering cross-section of H-nuclei gives neutrons important advantages for measuring dynamics in biological systems, and to date this has been achieved using standard neutron-wavelength (CW) or time-of-flight (TOF) techniques [5-20]. In an earlier publication [21] we have shown how new concepts of instrumentation, using either CW or TOF techniques can provide spectrometers specifically tailored to the needs of biophysics. We call these spectrometers elastic scattering spectrometers (ESS) because the energy spectrum is scanned by changing the energy resolution, rather than varying either the incident or scattered neutron-energy. In the current publication we examine the possibility of achieving this spectroscopy using neutron spin echo (NSE) [22], a technique normally associated with high resolution. However, there are disadvantages with NSE, in particular the need to polarise the incident beam, and the lower intensity for incoherent scattering (although this can be circumvented by deuteration). The central question is whether these are outweighed by the advantages of NSE: very high resolution and inherent focussing opportunities.

The underlying principle of all ESS methods is resolution elastic neutron scattering (RENS), for which we only give an outline here, the reader being referred to [21,23,24] for a more rigorous treatment. In the simplest case, the dynamics of the system under study has some characteristic relaxation time, $\tau_{SYSTEM}$, and in most biophysics studies this is the major interest. Considering the instrumental resolution in time, $\tau_{RES}$, if $\tau_{RES}$ is significantly shorter or longer than $\tau_{SYSTEM}$ then the measured elastic scattering intensity, $S_R(Q,\omega=0; \tau_{RES})$, has a weak dependence on $\tau_{RES}$. In the crucial intermediate region of resolution where $\tau_{RES}$ is comparable to $\tau_{SYSTEM}$ there will be a point of inflection in the plot of $S_R$ against $\tau_{RES}$ at the point where $\tau_{RES} = \tau_{SYSTEM}$ (Fig 1). Formally, the measured elastic-scattering intensity is the time integral of the product of the (system) intermediate scattering function with the (instrument) energy resolution function:

$$S_R(Q,\omega=0;\tau_{RES},\tau) = \int_{-\infty}^{\infty} R(t;\tau_{RES})I(Q,t;\tau)dt \quad (1)$$

Clearly, the detailed shape of the $S_R$ vs $\tau_{RES}$ contains information about the relaxation process(es), but discussion of this is beyond the scope of this paper. Paradoxically, the conception of an ESS spectrometer reduces to "what is the best way to worsen the instrumental resolution?". We emphasise that this paper examines the concept using NSE for ESS, and makes no attempt to propose a specific, or even a practical, design. For example, whilst any new NSE instrument would probably use zero-field spin-echo (ZFSE), for simplicity we will restrict our discussion to conventional fields and polarisation.

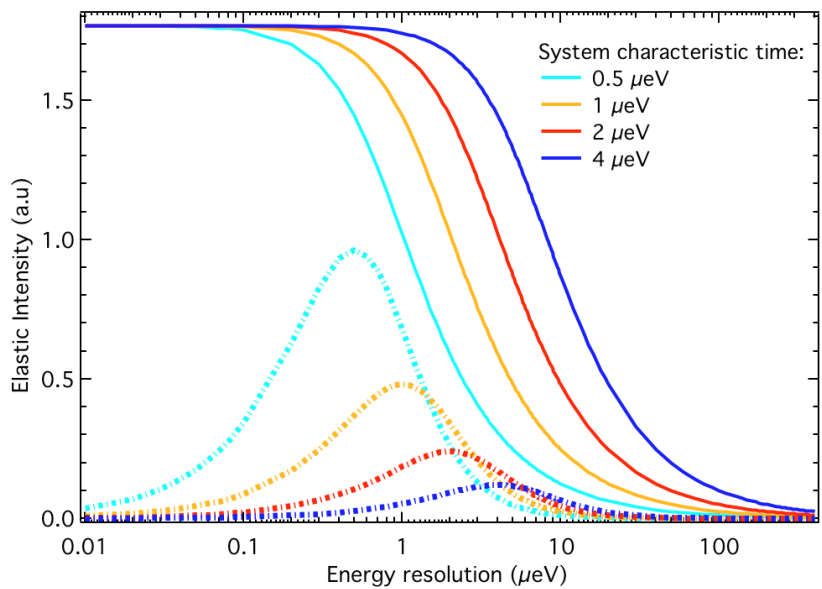

**FIGURE 1.** The ESS spectrum: elastic intensity as a function of the energy resolution. The inflection points occur where the energy resolution is equivalent to the system relaxation times, confirmed also by the second derivative (dashed lines) showing a maximum at the energy resolution values that match the system relaxation times.

## THE ESS SPIN-ECHO CONCEPT

An ESS spectrometer can be defined as a neutron instrument able to access the elastic-scattering intensity as a function of the energy resolution, as in Fig 1. In the measurement of dynamics there are usually 2 crucial parameters. Firstly, the length of the time window that can accessed, and secondly, the longest relaxation time that can be reached. Our previous designs for ESS spectrometers (CW and TOF) covered about 3 orders of magnitude in time, with the longest relaxation time being a few nanoseconds [21]. The incentive of using NSE is to improve on both of these, the central questions being: is this actually possible, and would it be worth examining in more detail at some later stage?

We will now outline the classical basics of NSE and how this would relate to an ESS spectrometer, considering first the layout of standard NSE (Fig. 2). We then consider focussing possibilities that would be required for the small sample-sizes that can be a crucial constraint for biological systems. Considering a single neutron with wavelength, $\lambda_{in}$, and velocity, $v_{in}$, polarized along $z$ passing a trajectory of length, $L$, over which there is a magnetic field of magnitude, *B*. The neutron spin experiences a torque from the magnetic field perpendicular to the spin direction and precesses with a Larmor frequency $\omega_L = \Upsilon B$, where $\Upsilon$ is the gyromagnetic ratio of the neutron (i.e. $1.832 \times 10^8$ s$^{-1}$T$^{-1}$) and *B* is the magnitude of the magnetic field. As a result, the total precession depends on the magnitude of *B*, and the time that the neutron spends in the field (i.e., the length of the neutron trajectory and its wavelength):

$$\varphi_1 = \gamma \frac{m}{h} B L \lambda_{in} \quad (2)$$

The neutron then interacts with the sample before passing through a second field that is identical to the first but with the field, *B*, reversed. Their polarisation is then analysed.

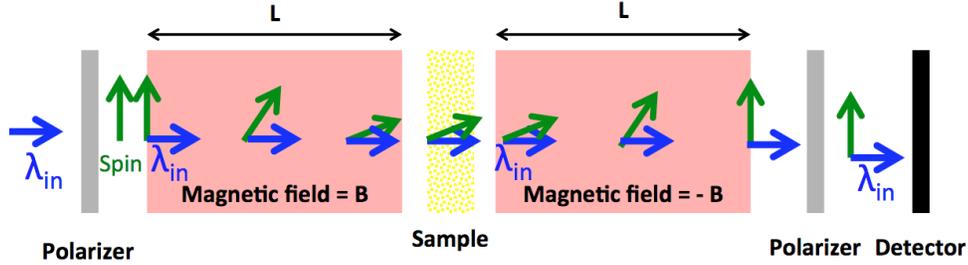

**FIGURE 2.** Basic layout of standard NSE concept in the echo condition, i.e. where there is no change in the neutron energy due to the scattering process with the sample.

In the ideal situation (i.e. perfect instrument) and with a perfect elastic-scatterer (e.g. vanadium) the precession of the spin in the first magnetic region is exactly unwound in the second and the net spin would be unchanged, i.e. the total precession angle is zero, $\Delta\varphi = \varphi_2 - \varphi_1 = 0$, that is "zero depolarisation". This is independent of the incident neutron wavelength. Clearly, in a real case, the sample is not a prefect elastic scatterer, and, in turn, it is able to change the wavelength of the neutrons during the scattering process so that the scattered neutron in the second field no longer "unwinds" the precession in the first field. The sensitivity of the polariser system $\varepsilon_{polariser}$ (that usually is few degrees) must be adequate to measure the overall depolarization.

$$\Delta\varphi = \varphi_2 - \varphi_1 = \gamma \frac{m}{h} \cdot B \cdot L \cdot \Delta\lambda > \varepsilon_{polariser} \quad (3)$$

If the resolution of the instrument is not able to resolve the system dynamics, then the system appears to be at rest, otherwise at higher resolution the instrument resolves the system dynamics. By Eq. (3) it is clear that for any change in the neutron wavelength (energy), $\Delta\lambda$, there is a lower limit for the magnitude of the magnetic field (for a fixed length) above which the instrument is able to resolve the dynamics:

$$B > \frac{h}{\gamma m} \frac{1}{L} \frac{1}{\Delta\lambda} \varepsilon_{polariser} \quad (4)$$

The higher the field (and its length), the better the instrumental resolution.

Our ESS spin-echo concept is to measure only the depolarization ratio as a function of changing the fields, before and after the sample, in a symmetric way such that we obtain the required plot of elastic scattering intensity against resolution, which is the ESS spectrum illustrated in Fig. 1. In practice there is a number of ways to change the resolution such as changing the wavelength range with a velocity-selector, and/or, changing the length for the fields (when using ZFSE). Clearly, combinations of all these would provide an instrument with a very wide dynamic range. We will discuss the use of shaped fields later in this paper.

Because we are only concerned with the ESS-NSE concept we will restrict our resolution-range discussion to the simplest parameters, wavelength and path integral (via *B* and *L*), accepting that other factors such as field homogeneity will also play a role. From Eq. (3) it is possible to estimate the total precession angle as a function of the energy transfer $\Delta E$:

$$\Delta\varphi = 10^0 \cdot B[T] \cdot L[m] \cdot \lambda^3[\text{\AA}^3] \cdot \Delta E[\mu eV] \quad (5)$$

By taking round estimates for the optimum values: $B=1T$; $L=1m$; $\lambda=10\text{\AA}$, we estimate the order of the best energy resolution achievable to be $\Delta E \approx 10^{-3} \mu eV$ (resolution time of $\tau_{RES} \approx ns$). To make the resolution worst we can reduce any combination of the field, the length, and the wavelength. Estimating reasonable values for these from existing instrumentation we obtain: $B=0.01T$; $L=0.1m$; $\lambda=3Å$, the corresponding worst resolution being $\Delta E \approx 10^2$ $\mu eV$, (resolution time a few of tens of picoseconds). As a result we estimate that ESS-NSE, could conceptually access 5 orders of magnitude in energy resolution, probing dynamical relaxations from nanoseconds to tens of picoseconds. There is a number of possibilities to extend this further but we will not discuss these here.

As anticipated, a simple ESS-NSE instrument concept shows a considerable extension to the time windows of the CW and TOF options [21], some of which overlaps the techniques, but more importantly extends to considerably longer times. However, the CW and TOF options were able to produce spatial focussing, which is an important advantage for biophysics studies. We now consider whether spatial focusing can also be achieved for ESS-NSE.

Simply adding a focussing monochromator (in addition, or place of, the velocity selector) to the existing concept will not work because, particularly for incoherent elastic scattering, there can be large differences between the path-lengths of the neutron in the two fields. Here, we examine the possibility of using shaped fields [25-28] to achieve an echo for all elastically scattered wavelengths and trajectories. This requires that the product of total field-integral (first plus second fields) and wavelength be the same regardless of wavelength. The concept is illustrated in the Fig. 3. The focussing monochromator provides not only focussing, but also a spatial separation of wavelengths, providing the opportunity to balance the differing precessions associated with these by using shaped magnetic fields. By making the first field nominally triangular in shape, we can arrange that the shorter wavelengths from the monochromator segments pass through a longer field (higher field-integral) than those with longer wavelengths. Neutrons scattered by the sample all pass the same standard field to the analyser as in normal NSE. Note that the precession is not focussed at the sample, but at the detector, so at the sample, shorter wavelengths are "ahead" of longer wavelengths in terms of precession. This is broadly analogous to time focusing where the neutron arrival-time is focussed at the detectors (not the sample) for optimal performance in the elastic-scattering region. In this way the uncertainties in $B$, $L$ and wavelength need not be significantly different from in the normal NSE arrangement, but clearly we have sacrificed some incident flux by using the monochromator in place of (or in addition to) a velocity selector.

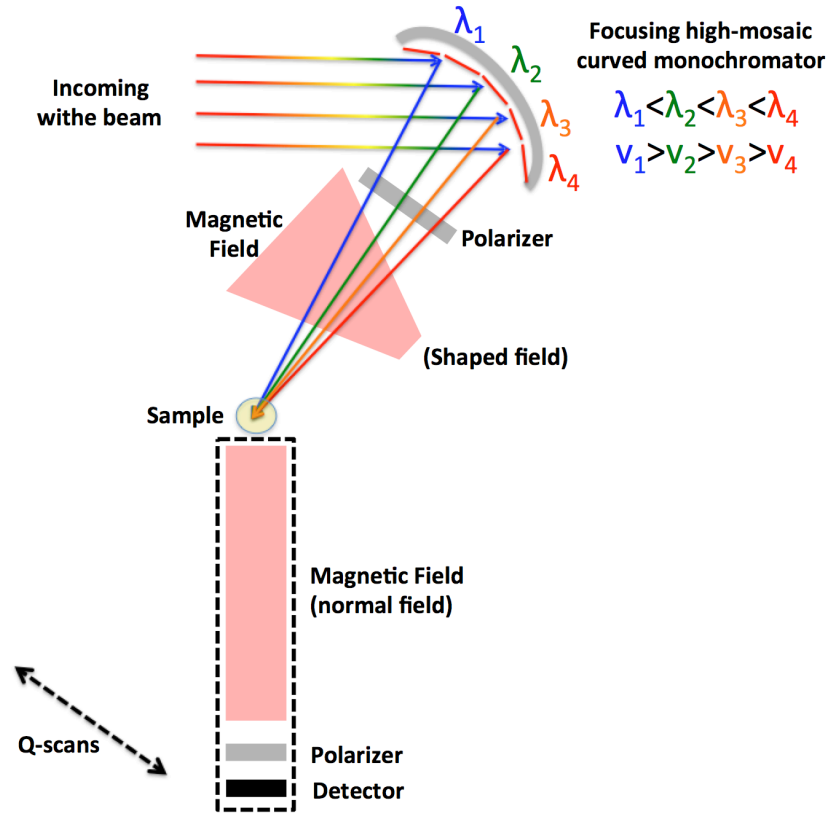

**FIGURE 3.** Illustration of focussing using a curved monochromator whilst retaining a common echo at the detector for all elastically scattered wavelengths. Sizes and shapes of the monochromator and fields are schematic only.

# CONCLUSIONS

This paper has built on our previous work [21] which demonstrated that dynamical information can be conveniently accessed by measuring the elastic intensity only as a function of the instrumental energy resolution, i.e. ESS. The advantage, particularly for biophysics, is that this overall relaxation time is obtained in a parameterless way, regardless of the complexity of the system. Here we wanted to examine an ESS spin-echo concept, taking advantage of the inherent high resolution and broad time-range of spin echo. We contend that simply altering the resolution by changes to the field strength, length or wavelength range would provide changes in resolution from nanoseconds to tens of picoseconds, and that an existing spin-echo set-up on a test-beam could be used to verify the concept.

Because we only measure elastic scattering the overall data rate should be good, but with the emphasis on biophysics we have examined increasing the beam-current to small samples by spatial focussing. By using a curved monochromator that provides spatial focussing and spatial separation of wavelengths, we have presented a concept that would not only focus the beam on the sample, but using a shaped primary field. Length and wavelength uncertainties are at least partially removed from the resolution. This idea would have to be tested by simulation because it is unlikely that the required elements already exist.

In our previous work on ESS using CW and TOF methods we learnt that we can only try to be quantitative about performance and design by making a full instrument simulation. Because we have yet to make this simulation for the current ESS-NSE version we have made no attempt to be quantitative in the present work, restricting ourselves to the instrumental concept and some reasonable figures. There is a host of alternatives and elaborations that are now available within the spin-echo technique, but we have avoided complicating our main concept by introducing these.

We hope that this new option for ESS stimulate discussion within (and between) the biophysics and neutron instrumentation groups that would be of benefit to the whole community.

# ACKNOWLEDGMENTS

A.B. acknowledges support from (i) the European Community under the Marie-Curie Fellowship Grants HYDRA (No. 301463) and PSI-FELLOW (No. 290605), and from (ii) Science Foundation Ireland (SFI) under the Start Investigator Research Grant 15-SIRG-3538, with additional support provided by the School of Physics, University College Dublin, Ireland, and the Laboratory for Neutron Scattering, Paul Scherrer Institute, Switzerland. A.B. acknowledges also support from the Australian Government under the 2012 Endeavour Research Fellowship (No. 3341-2012).